\documentstyle[12pt]{article}

\begin{document}

\centerline{\Large A lattice quantum gravity model with surface-like} 
\centerline{\Large excitations in 4-dimensional spacetime}
\vspace{0.5cm}
\centerline{\large\it Junichi Iwasaki}
\vspace{0.3cm}
\centerline
{Physics Department, University of Pittsburgh, Pittsburgh, PA 15260, USA}
\vspace{0.2cm}
\centerline{E-mail: iwasaki@phyast.pitt.edu}
\vspace{0.3cm}
\centerline{June 24, 2000}


\begin{abstract}
A lattice quantum gravity model in 4 dimensional Riemannian spacetime 
is constructed based on the SU(2) Ashtekar formulation of general relativity.
This model can be understood as one of the family of models sometimes called
``spin foam models.''
A version of the action of general relativity in continuum is introduced
and its lattice version is defined.
A dimensionless ``(inverse) coupling'' constant is defined so that
the value of the action of the model is finite per lattice point.
The path integral of the model
is expanded in the characters and shown to be written 
as a sum over surface-like excitations in spacetime.
A 3 dimensional version of the model exists and can be
reduced to lattice BF theory.
The expectation values of some quantities are computed in 3 dimensions
and the meanings of the results are discussed.
Although the model is studied on a hyper cubic lattice for simplicity, 
it can be generalized to a randomly triangulated lattice 
with small modifications.
\end{abstract}


\section{Introduction}\label{sec:introduction}

A surface theoretic view of
non-perturbative quantum gravity was studied 
some time ago based on 3 dimensional general relativity as 
a reformulation of the Ponzano-Regge model in Riemannian spacetime 
\cite{JI3D}.
Since 3 dimensional general relativity has no local degrees of freedom,
the Ponzano-Regge model is known to be a topological field theory
called lattice BF theory in 3 dimensions.
A 4 dimensional generalization was made by Ooguri and is known as
lattice BF theory in 4 dimensions, also a topological field theory
\cite{HO}.
This Ooguri model also allows the surface theoretic view.
However, since 4 dimensional general relativity contains 
local degrees of freedom,
the Ooguri model is not a quantization of general relativity.

In order to pursue 4 dimensional generalizations of 
the surface theoretic view of the Ponzano-Regge model with local degrees of
freedom, some investigations have been being made. 
They can be divided roughly to four strategies.
One \cite{CR} is the use of canonical theory called loop quantum gravity 
\cite{review}
based on Ashtekar formulation of general relativity \cite{AA}.
It uses a Hamiltonian of the canonical theory to compute quantum amplitudes.
In this strategy,
the definition of ``correct'' Hamiltonian is still uncertain
although the canonical theory is able to provide physical interpretations.
Another \cite{MR,FK,JI4D}
is the use of the Plebanski action of general relativity 
in Reimannian spacetime. 
This action has the form of BF theory with additional constraint terms 
\cite{JP}.
It computes a path integral of the action.
In this strategy, a well regulated computation of the constraint terms 
is still missing and, in addition, the action functional is available only
in Reimannian spacetime.
Another \cite{BC,JB}
is the use of ``quantized'' lattice BF theory and additional ``quantum level'' 
constraint conditions.
In this strategy, the formulation is strictly mathematical and it necessitates
help from other theories for physical interpretations.
The other \cite{MS} is based on physical reasoning such as causality
and statistical criticality.
This strategy wants a systematic formulation.

In the present work, a fifth strategy is applied to construct 
a surface theoretic model for 4 dimensional quantum gravity.
We introduce a version of the action of general relativity in continuum.
This action is derived from the action proposed some time ago by
Samuel and by Jacobson and Smolin \cite{SSJ}.
The Ashtekar formulation can be constructed from the latter.
We define a lattice version of this action and its path integral.
The variables integrated are closely related to the SU(2) (real valued)
variables of the Ashtekar formulation.
However, the model avoids the use of Hamiltonian of canonical theory,
which is a main difficulty so far for quantization.
The spacetime the model assumes has signature $(+1,+1,+1,+1)$,
Riemannian spacetime.
In order to make the lattice action finite per lattice point, 
a dimensionless ``(inverse) coupling'' constant is defined in analogy to 
Wilson's formulation of lattice gauge theory \cite{KW}.
The finiteness of the action with finite lattice allows the path integral
without gauge fixing finite.
The path integral of the model is computed as a character expansion 
and shown to be written as a sum over surface-like excitations in 
spacetime.

We show that a 3 dimensional version of the model exists and can be
reduced to lattice BF theory.
In this sense, the model is a 4 dimensional generalization of 
the Ponzano-Regge model with local degrees of freedom.
The expectation values of four quantities are computed in 3 dimensions.
Two of them are basic variables of the theory and are SU(2) gauge
dependent quantities.
Another is the Wilson loop. 
It is SU(2) gauge independent but diffeomorphism dependent.
The other is the action itself.
It is invariant under not only SU(2) gauge transformations but
also diffeomorphisms.
The first three are shown to be always zero unless an appropriate gauge 
is fixed. 
The other is zero if the (inverse) coupling constant is brought to
an appropriate limit.
These are very simple results but important checks for consistency.
We discuss the meaning of these results.
Note that this kind of computations have not been performed in
the other strategies while the computation of the path integral
(or sometimes called transition amplitude, projector or
partition function) is always focussed on in all the strategies.

The model is defined on a fixed hyper cubic lattice for simplicity.
However, the construction can be generalized to a randomly
triangulated lattice with small modifications.
{}Further more, the sum over different triangulations can be done,
if necessary, by applying a technology developed in a recent work
\cite{sumover}.

This paper is organized as follows.
In Sec. \ref{sec:action}, we restructure the action functional of
general relativity so that it can be utilized to define the model.
In Sec. \ref{sec:model}, the model is defined and 
its path integral is studied.
In Sec. \ref{sec:3dim}, three dimensional version of the model is studied.
In Sec. \ref{sec:conclusion}, we conclude the work.
In Appendix \ref{sec:useful}, mathematical formulae 
utilized in this work are listed.


\section{The action of general relativity}\label{sec:action}

The action we make use of is the one proposed by Samuel and by
Jacobson and Smolin \cite{SSJ}.
It is the Palatini action but an additional term.
The additional term is a topological term and
not responsible to the local degrees of freedom
of general relativity.
{}From the Palatini action, one can construct a canonical theory
by splitting space and time.
However, it contains constraints of the second class in addition to
those of the first class in the sense of Dirac.
Therefore, the constrained phase space is not well defined unless 
the second class constraints are explicitly solved.

On the other hand, from the present action, one can construct
Ashtekar's canonical theory, which contains only constrains of 
the first class.
It means that the constrained phase space is well defined in the sense
that it is invariant under transformations produced by Hamiltonian
and diffeomorphism generators.
The difference between the two canonical theories is due to the canonical
transformation produced by the topological term in the 
Samuel-Jacobson-Smolin action.
We restructure this action for the use in the following section.

Let us introduce variables we need.
We use tetrad field $e_\mu^I$ rather than metric $g_{\mu\nu}$.
They are related with each other by $g_{\mu\nu}=e_\mu^Ie_\nu^J\eta_{IJ}$.
Here $\eta_{IJ}$ is the Euclidean or Minkowskian metric of
``internal'' spacetime depending on the value of $\sigma$ with
signature $(\sigma,+1,+1,+1)$.
The capital alphabets $I,J\cdots$ are used for internal spacetime
indices $\{0,1,2,3\}$ and the Greek letters are for spacetime
indices $\{0,1,2,3\}$.
Below, we also use the lower case alphabets $i,j\cdots$ for the space
componets of the internal spacetime indices $\{1,2,3\}$
and for the adjoint indices of SU(2) group.

The so-called spin-connection $\omega_\mu^{IJ}$ is written in terms of
the tetrad field as follows.
\begin{eqnarray}
&\omega_\mu^{IJ}&:=2e^{\sigma [I}\partial_{[\mu}e_{\sigma]}^{J]}
-e^{\sigma I}e^{\nu J}e_{\mu K}\partial_{[\sigma}e_{\nu]}^K
\nonumber\\&&
=\sigma\epsilon^{\sigma\alpha\beta\gamma}\epsilon^{IJ}_{\ \ MN}
e_\alpha^Me_\beta^N\left(e_\gamma^K\partial_{[\mu}e_{\sigma]K}
+{1\over2}e_\mu^K\partial_\gamma e_{\sigma K}
\right).
\end{eqnarray}
Here $e^{\mu I}$ is the inverse tetrad.
The Riemann tensor $R_{\mu\nu}^{IJ}$ is written in terms of 
the spin-connection as follows.
\begin{eqnarray}
&&
R_{\mu\nu}^{IJ}:=2\partial_{[\mu}\omega_{\nu]}^{IJ}
+2\omega_{[\mu}^{IK}\omega_{\nu] K}^{\ \ \ J}.
\end{eqnarray}
In terms of these traditional variables, we define connection and 
curvature variables $A_\mu^{IJ}$ and $F_{\mu\nu}^{IJ}$ as follows.
\begin{eqnarray}
&&
A_\mu^{IJ}:=
\omega_\mu^{IJ}+{1\over2}\gamma\epsilon^{IJ}_{\ \ KL}\omega_\mu^{KL} 
={1\over2}\gamma\epsilon^{IJ}_{\ \ KL}A_\mu^{KL}+
(1-\gamma^2\sigma) \omega_\mu^{IJ},
\\&&
{}F_{\mu\nu}^{IJ}:=R_{\mu\nu}^{IJ}+
{1\over2}\gamma\epsilon^{IJ}_{\ \ KL}R_{\mu\nu}^{KL}
={1\over2}\gamma\epsilon^{IJ}_{\ \ KL}F_{\mu\nu}^{KL}+
(1-\gamma^2\sigma) R_{\mu\nu}^{IJ},
\\&&
{}F_{\mu\nu}^{IJ}=2\partial_{[\mu} A_{\nu]}^{IJ}
+A_{[\mu}^{IK}A_{\nu]K}^{\ \ \ J}
+(1-\gamma^2\sigma)\omega_{[\mu}^{IK}\omega_{\nu]K}^{\ \ \ J}.
\end{eqnarray}
Here $\gamma$ is a parameter and chosen to be $\gamma=1$ to
relate the model with SU(2) (real valued) Ashtekar formulation.
If one chooses the value such that $\gamma^2\sigma=1$, then
the variables $A_\mu^{IJ}$ and $F_{\mu\nu}^{IJ}$ are self-dual
with respect to the internal spacetime indices.
We keep $\gamma$ unspecified in the course of development 
within this section.

Among the components of $A_\mu^{IJ}$ and $F_{\mu\nu}^{IJ}$,
we use specifically $A_\mu^{0i}$ and $F_{\mu\nu}^{0i}$.
$F_{\mu\nu}^{0i}$ is written in terms of $A_\mu^{0i}$ and
$\omega_\mu^{IJ}$ as follows.
\begin{eqnarray}
&&
{}F_{\mu\nu}^{0i}=2\partial_{[\mu} A_{\nu]}^{0i}
-\sigma\gamma\epsilon^{0i}_{\ \ jk}A_{\mu}^{0j}A_{\nu}^{0k}
+(1-\gamma^2\sigma)\left[2\omega_{[\mu}^{0k}\omega_{\nu]k}^{\ \ \ i}
+{1\over2}\gamma\epsilon^{0k}_{\ \ mn}
\omega_{[\mu}^{mn}\omega_{\nu]k}^{\ \ \ i} \right].
\end{eqnarray}
The pull-back of $A_\mu^{0i}$ to foliated space of spacetime is precisely
the SU(2) (real valued) connection variables of the Ashtekar
canonical formulation.

Let us explicitly write the action.
{}First, the Palatini action derived from the Hilbert-Einstein action is
\begin{eqnarray}
&S_0[e,\omega]&:=
{1\over l^2_p}\int d^4x \sqrt{\sigma g}R
={1\over l^2_p}\int d^4x ee_I^\mu e_J^\nu R_{\mu\nu}^{IJ}
\nonumber\\&&
={1\over 2l^2_p}\int d^4x \epsilon^{\mu\nu\lambda\sigma}\epsilon_{IJKL}
e^I_\mu e^J_\nu R_{\lambda\sigma}^{KL}.
\end{eqnarray}
Here,
$g$ is the determinant of metric $g_{\mu\nu}$, $R$ is the scalar curvature,
$l_p$ is the Planck length constant, $e$ is the determinant of $e_\mu^I$
and $\epsilon^{\mu\nu\lambda\sigma}$ is the alternating density.
By adding a topological term as mentioned above, the action we make use
of is
\begin{eqnarray}
&S[e,A]&:=
{1\over\gamma l^2_p}\int d^4x \epsilon^{\mu\nu\lambda\sigma}
e_{\mu I}e_{\nu J}\left[{1\over2}\gamma\epsilon^{IJ}_{\ \ KL} 
R_{\lambda\sigma}^{KL}+R_{\lambda\sigma}^{IJ}\right]
\nonumber\\&&
={1\over\gamma l^2_p}\int d^4x \epsilon^{\mu\nu\lambda\sigma}
e_{\mu I}e_{\nu J}F_{\lambda\sigma}^{IJ}.
\end{eqnarray}
This is the Samuel-Jacobson-Smolin action.
By dividing space and time indices for internal spacetime
and writing in terms of $e_\mu^0$, $e_\mu^i$ and $A_\mu^{0i}$,
the action becomes
\begin{eqnarray}
&&
S[e,A]
={1\over\gamma l^2_p}\int d^4x \epsilon^{\mu\nu\lambda\sigma}
\left[2e_{\mu 0}e_{\nu i}F_{\lambda\sigma}^{0i}
+e_{\mu i}e_{\nu j}F_{\lambda\sigma}^{ij}\right]
\nonumber\\&&
={1\over\gamma l^2_p}\int d^4x \epsilon^{\mu\nu\lambda\sigma}
\left[2e_{\mu 0}e_{\nu i}F_{\lambda\sigma}^{0i}
+e_{\mu i}e_{\nu j}
\left[\gamma\epsilon^{ij}_{\ \ 0k}F_{\lambda\sigma}^{0k}+
(1-\gamma^2\sigma)R^{ij}_{\lambda\sigma}\right]\right]
\nonumber\\&&
={1\over\gamma l^2_p}\int d^4x \epsilon^{\mu\nu\lambda\sigma}
\left[
\sigma\left(2e_{[\mu}^0e_{\nu]i}
+\gamma\epsilon^{0}_{\ ijk}e_\mu^je_\nu^k\right) F_{\lambda\sigma}^{0i}
+(1-\gamma^2\sigma)e_{\mu i}e_{\nu j}R^{ij}_{\lambda\sigma}
\right]
\nonumber\\&&
={1\over\gamma l^2_p}\int d^4x \epsilon^{\mu\nu\lambda\sigma}
\left\{
\sigma\left(2e_{[\mu}^0e_{\nu] i}
+\gamma\epsilon^{0}_{\ ijk}e_\mu^je_\nu^k\right) 
\left(2\partial_{[\lambda}A_{\sigma]}^{0i}
-\sigma\gamma\epsilon^{0i}_{\ \ jk}
A_\lambda^{0j}A_\sigma^{0k}\right)
\right.\nonumber\\&&
+(1-\gamma^2\sigma)
\left[\sigma\left(2e_{[\mu}^0e_{\nu] i}
+\gamma\epsilon^{0}_{\ ijk}e_\mu^je_\nu^k\right) 
\left(2\omega_{[\lambda}^{0k}\omega_{\sigma]k}^{\ \ \ i}+{1\over2}
\gamma\epsilon^{0k}_{\ \ mn}
\omega_{[\lambda}^{mn}\omega_{\sigma]k}^{\ \ \ i}
\right)\right.
\nonumber\\&&\left.\left.
+e_{\mu i}e_{\nu j}R^{ij}_{\lambda\sigma}\right]\right\}.
\label{eq:action}
\end{eqnarray}
Here, we consider $\omega_\mu^{IJ}$ and $R_{\mu\nu}^{ij}$ as
functions of $e_\mu^0$ and $e_\mu^i$.
Note that if $\gamma^2\sigma=1$, then the terms proportional to
$(1-\gamma^2\sigma)$ go away and the action gets simplified.
We will utilize the case that $\gamma=1$ and $\sigma=1$, 
Riemannian spacetime.


\section{The model}\label{sec:model}

\subsection{The action}

The action discussed in the previous section
for Riemannian spacetime $(\sigma=+1)$ with $\gamma=1$ is 
\begin{eqnarray}
&&
S_R[e,A]=
{1\over l^2_p}\int d^4x \epsilon^{\mu\nu\lambda\sigma}
\left(2e_{[\mu}^0e_{\nu] i}
+\epsilon^{0}_{\ ijk}e_\mu^je_\nu^k\right) 
\left(2\partial_{[\lambda}A_{\sigma]}^{i}
-\epsilon^{0i}_{\ \ jk}
A_\lambda^{j}A_\sigma^{k}\right).
\end{eqnarray}
Hereafter, we write $A_\mu^i$ for $A_\mu^{0i}$.
We note that if 
$2e_{[\mu}^0e_{\nu]}^i+\epsilon^{0i}_{\ \ jk}e_\mu^je_\nu^k$
is replaced by a 2-form variable denoted by $B_{\mu\nu}^i$
with the condition that 
$\epsilon^{\mu\nu\lambda\sigma}B_{\mu\nu}^iB_{\lambda\sigma}^j$
is positionwisely proportional to $\delta^{ij}$, then
the resulting action is precisely the Plebanski action.
This additional condition is supposed to restore
$B_{\mu\nu}^i=2e_{[\mu}^0e_{\nu]}^i+\epsilon^{0i}_{\ \ jk}e_\mu^je_\nu^k$.
Ways of constructing a surface theoretic model 
starting from the Plebanski action have been explored.
The condition is usually imposed with Lagrange multipliers.
However, it has turned out that the inclusion of the condition
in terms of Lagrange multilpiers is still a difficult task.
In the present work, we do not follow any of those ways.
Instead, we define a corresponding action directly on the lattice
without introducing $B_{\mu\nu}^i$ and utilizing Lagrange multipliers.

In order to define the model, 
we rewrite the action in a slightly compact fashion so that
the construction of the lattice version is straightforward.
An idea is the following.
The tetrad  variables (or their modifications) are often considered 
su(2) algebra valued.
{}For example, in the Ashtekar formulation, the pull-back of the (inverse)
tetrads to space  have su(2) algebra indices and , 
in the Plebanski formulation, the 2-form variable $B_{\mu\nu}^i$ are
su(2) algebra valued.
However, we do not consider here the tetrad variables su(2) algebra
valued but do consider them SU(2) group valued.
If one defines $e_\mu:=e_\mu^0+i\sigma_ie_\mu^i$ with the Pauli
matrices $\sigma_i$ $(i=1,2,3)$, then the complicated expression
$2e_{[\mu}^0e_{\nu]}^i+\epsilon^{0i}_{\ \ jk}e_\mu^je_\nu^k$
becomes just  the non-trace part of $e_\mu^\dagger e_\nu$.
If $e_\mu$ is normalized so that 
$e_\mu^0e_\mu^0+e_\mu^1e_\mu^1+e_\mu^2e_\mu^2+e_\mu^3e_\mu^3=1$,
then it is an SU(2) group element.
This small change of view dramatically simplifies the construction of 
the model and  makes the extension to other dimensional spacetimes 
straightforward.

Before introducing the lattice, let us count the number of degrees of
freedom of the variables. 
$e_\mu^I$ and $A_\mu^i$ have respectively 16 and 12 degrees of freedom 
per spacetime point.
{}From thses variables, if one constructs canonical pairs of variables,
one finds 2 physical degrees of freedom from $e_\mu^I$ and 2 from $A_\mu^i$.
The other degrees of freedoms are for the constraints and gauge trajectries 
in addition to the freedom of specifying a spacetime foliation.
Let us be slightly more concrete.
Out of 16 degrees of freedom of $e_\mu^I$, 3 fix a spacetime foliation and
4 are used for Lagrange multipliers imposing three diffeomorphism and
a Hamiltonian constraints.
Out of 12 degrees of freedom of $A_\mu^i$, 3 are used for Lagrange multipliers
imposing three Gauss gauge constraints.
Hence, 9 degrees of freedom in each are left.
These are exactly the number of degrees of freedom of the Ashtekar canonical
theory.
Then by subtracting 7 for the constraints and 7 for the gauge 
degrees of freedom, one finds the correct number of physical degrees of
freedom.


\subsection{Lattice}\label{subsec:lattice}

We fix a lattice.
We do not take sum over different lattices.
In general, the model can be defined on a randomly triangulated lattice.
However, technically it is difficult, if not impossible, to compute
the quantum amplitudes analytically or even numerically
if the structure of lattice varies from place to place.
{}For this reason, we use a hyper cubic lattice.
{}First we introduce a pair of hyper cubic lattices.
One is dual to the other.
Denote them $\Delta$ and $\Delta^*$ respectively.
Then we further restrict ourselves to the case that $\Delta^*$ is $\Delta$
itself.
This is possible because the dual lattice of a hyper cubic lattice is
a hyper cubic lattice.
This restriction dramatically simplifies the construction of the model
and helps computations in practice.

The lattices $\Delta$ and $\Delta^*$ are
a pair of 4-dimensional hyper cubic lattices dual to each other.
They consist of
0-cells (vertices), 1-cells (edges), 2-cells (square faces), 
3-cells (cubes) and 4-cells (hyper cubes).
Every k-cell of one lattice intersects with a (4-k)-cell of the other
lattice at a point inside the cells.
Every cell of the lattices has one and only one intersection.
The correspondence of k-cells of $\Delta$ and (4-k)-cells of $\Delta^*$
is one-to-one.
The corresponding k-cell of $\Delta$ and (4-k)-cell of $\Delta^*$ are
called dual to each other.
There is a limit at which every intersection coincides with 
one of the vertices so that the two lattices are identical 
$\Delta=\Delta^*$.

Let us be more specific.
{}Fix $\Delta$ and $\Delta^*$.
Both have a lattice spacing $\varepsilon$.
One is the displacement of the other by the distance $\varepsilon/2$
in all the four directions.
Let $x$ denote (the position of) a vertex of $\Delta$.
The (positions of) adjacent vertices are $x\pm\varepsilon^\mu$
with $\mu=0,1,2,3$, where $\varepsilon^\mu$ is the increment in the 
$\mu$-th direction with the magnitude $\varepsilon$.
Note that $x\pm\varepsilon^0$, for example, is a short hand notation for
$(x^0\pm\varepsilon,x^1,x^2,x^3)$.
Let $x^*$ denote (the position) of a vertex of $\Delta^*$ at
$x+\varepsilon^0/2+\varepsilon^1/2+\varepsilon^2/2+\varepsilon^3/2$.
The face of $\Delta$ specified by (the positions of) four vertices 
$x$, $x+\varepsilon^\mu$, $x+\varepsilon^\nu$ and 
$x+\varepsilon^\mu+\varepsilon^\nu$ is dual to
the face of $\Delta^*$ specified by (the positions of) four vertices 
$x^*$, $x^*-\varepsilon^\lambda$, $x^*-\varepsilon^\sigma$ and 
$x^*-\varepsilon^\lambda-\varepsilon^\sigma$, where the indices
are chosen such that $\epsilon^{\mu\nu\lambda\sigma}=1$.
There are six pairs of faces for given $x$ up to the orientation of 
the face.
These dual faces are important in our construction of the model.
After taking the limit $\Delta^*\to\Delta$ such that $x^*$ coincides
with $x$, the dual faces are still clearly defined on the single lattice
$\Delta$.
The intersection of a dual pair of faces, one basing at $x$ and 
the other at $x^*$, coincides with $x=x^*$ at the limit.
Notice that the face basing at $x^*$ and in the $\mu\nu$ plane does not 
coincide with the face basing at $x$ and in the same plane at the limit.
One is in the $\mu$ and $\nu$ directions while the other is 
in the $-\mu$ and $-\nu$ directions from the coincident base point
$x=x^*$.
We will define our model on the lattice $\Delta (=\Delta^*)$.


\subsection{The lattice action}\label{subsec:latticeaction}

We define an action on the lattice $\Delta$ so that
it converges to the action in continuum as the lattice spacing
goes to zero.
{}First, let us define variables on the lattice as follows.
{}For an edge basing at $x$ and pointing at the $\mu$-th direction, 
define
\begin{eqnarray} 
&&
\zeta_\mu(x):=\zeta_\mu^0(x)+i\sigma_i \zeta_\mu^i(x):=
\exp[i\varepsilon A_\mu^i(x)\sigma_i/2],
\label{eq:zeta}
\\&&
\eta_\mu(x):=\eta_\mu^0(x)+i\sigma_i \eta_\mu^i(x):=
(\beta^{1/2}l_p)^{-1}\varepsilon[e_\mu^0(x)+ie_\mu^i(x)\sigma_i]
/\rho_\mu(x),
\label{eq:eta}
\\&&
\rho_\mu(x):=(\beta^{1/2}l_p)^{-1}\varepsilon
\sqrt{|e_\mu^0(x)|^2+|e_\mu^1(x)|^2+|e_\mu^2(x)|^2+|e_\mu^3(x)|^2}.
\label{eq:rho}
\end{eqnarray}
Here, $\varepsilon$ is the lattice spacing and
$\sigma_i$'s are the Pauli matrices, that is,
\begin{eqnarray}
&&
\sigma_1:=\left[\matrix{0&1\cr 1&0}\right],{\ \ }
\sigma_2:=\left[\matrix{0&-i\cr i&0}\right],{\ \ }
\sigma_3:=\left[\matrix{1&0\cr 0&-1}\right].
\end{eqnarray}
Note that the inside the square root in $\rho_\mu$ is equal to 
the metric element $g_{\mu\mu}$.
$\beta$ is the upper bound of 
$l_p^{-2}\varepsilon^2g_{\mu\mu} (\mu=0,1,2,3)$ and introduced
to regularize the path integral defined below.
$\rho_\mu$ is non-negative and less than $1$,
and $\zeta_\mu$ and $\eta_\mu$ are 
SU(2) matrices normalized such that
$|\zeta_\mu^0|^2+|\zeta_\mu^1|^2+|\zeta_\mu^2|^2+|\zeta_\mu^3|^2=1$ and
$|\eta_\mu^0|^2+|\eta_\mu^1|^2+|\eta_\mu^2|^2+|\eta_\mu^3|^2=1$.
{}For a face basing at $x$ and enclosed by four edges in the
$\mu$ and $\nu$-th direstions, define
\begin{eqnarray}
&&
U_{\mu\nu}(x):=\zeta_\mu(x)\zeta_\nu(x+\varepsilon^\mu)
\zeta_\mu^\dagger(x+\varepsilon^\nu)\zeta_\nu^\dagger(x),
\end{eqnarray}
without summing over $\mu$ and $\nu$.
Here, $\varepsilon^\mu$ is the increment in the $\mu$-th direction
with the magnitude $\varepsilon$.

The action on the lattice $\Delta$ 
is then defined as follows.
\begin{eqnarray}
&&
S_{\Delta}[\zeta,\eta,\rho]:=
-\beta\sum_{x\in\Delta}
\epsilon^{\mu\nu\lambda\sigma}\rho_\mu(x)\rho_\nu(x)
{\rm Tr}[\eta_\mu^\dagger(x)\eta_\nu(x)
U_{\lambda\sigma}(x)].
\end{eqnarray} 
Here, the sums over $\mu$, $\nu$, $\lambda$ and $\sigma$ have been
performed.
This lattice action is finite on the lattice $\Delta$ 
with $\varepsilon$ and $\beta$ fixed if the lattice size 
is finite.
This is because $\zeta_\mu$ and $\eta_\mu$ are compact SU(2) variables
and $\rho_\mu$ has lower and upper bounds in addition to the fact that
the number of degrees of freedom is finite on the lattice 
if the lattice size is finite.
The finiteness of the action lets the path integral without gauge
fixing non-divergent. 
In other words, the path integral along a gauge trajectory produces
just a finite overall multiplicative constant.
We call $\beta$ ``inverse coupling'' in analogy to Wilson's
formulation of lattice gauge theory.

Let us examine if the lattice action really converges to the continuum
action as the lattice spacing $\varepsilon$ goes to zero.
Take the limit $\varepsilon\to 0$.
In the limiting process, it can be shown that
\begin{eqnarray}
&&
\beta\rho_\mu(x)\rho_\nu(x)
\eta_{[\mu}^\dagger(x)\eta_{\nu]}(x)=
l_p^{-2}\varepsilon^2i\sigma_i\left(
2e_{[\mu}^0(x)e_{\nu]}^i(x)+\epsilon^{0i}_{\ \ jk}e_\mu^j(x)e_\nu^k(x)
\right),
\\&&
U_{[\lambda\sigma]}(x)\to{1\over2}\varepsilon^2i\sigma_i
\left(2\partial_{[\lambda}A_{\sigma]}^i(x)-\epsilon^{0i}_{\ \ jk}
A_\lambda^j(x)A_\sigma^k(x)\right)+{\cal O}(\varepsilon^3).
\end{eqnarray}
Therefore, the lattice action converges to
\begin{eqnarray}
&
S_{\Delta}[\zeta,\eta,\rho]\to&
l_p^{-2}\sum_{x\in\Delta}\varepsilon^4\epsilon^{\mu\nu\lambda\sigma}
\left(2e_\mu^0(x)e_{\nu i}(x)+\epsilon^{0}_{\ ijk}e_\mu^j(x)e_\nu^k(x)\right)
\times\nonumber\\&&
\left(2\partial_\lambda A_\sigma^i(x)-\epsilon^{0i}_{\ \ jk}
A_\lambda^j(x)A_\sigma^k(x)\right)
+{\cal O}(\varepsilon^5).
\end{eqnarray}


\subsection{The lattice path integral}\label{subsec:pathintegral}

The exponential of the action can be considered as (an extension of)
the graph-cylindrical function discussed in \cite{AL}.
The integral measure we use for the variable $A_\mu^i$ is
the Ashtekar-Lewandowski measure \cite{AL}.
The measure for SU(2)-compactified part of $e_\mu^0$ and $e_\mu^i$ is
analogously defined and that for the other part of them is the one
discussed in \cite{CR}.
The path integral has the form of
\begin{eqnarray}
&&
\int d\mu(A)d\mu(e)\Psi_{\Delta,\psi}(A,e):=
\int \prod_{x\in \Delta}\prod_{\mu}
d\zeta_\mu(x) d\eta_\mu(x) d\rho_\mu^4(x)
\times\nonumber\\&&
\psi(\zeta(x_1,A),\cdots\zeta(x_l,A);\eta(x_1,e),\cdots\eta(x_m,e);
\rho(x_1,e),\cdots\rho(x_n,e)),
\end{eqnarray}
where $\Psi_{\Delta,\psi}$ is a cylindrical function defined on
the lattice $\Delta$ in terms of 
a complex valued integrable function 
$\psi$ on $[SU(2)]^l\times[SU(2)]^m\times [0,1]^n$.
$d\zeta_\mu$ and $d\eta_\mu$ are the Haar measures for SU(2) elements
$\zeta_\mu$ and $\eta_\mu$ respectively.
$d\rho_\mu^4=4d\rho_\mu\rho_\mu^3$ is the radial integral of 
the 4-dimensional unit sphere.
$\rho_\mu$ is integrated from $0$ to $1$.

The path integral we will compute in the following section
is defined as follows.
\begin{eqnarray}
&&
Z_{\Delta}:=
\int d\zeta d\eta d\rho^4 
e^{-iS_{\Delta}[\zeta,\eta,\rho]}
\nonumber\\&&
=\int d\zeta d\eta d\rho^4 \prod_{x\in\Delta}
e^{i\beta\epsilon^{\mu\nu\lambda\sigma}\rho_\mu(x)\rho_\nu(x)
{\rm Tr}[\eta_\mu^\dagger(x)\eta_\nu(x)
U_{\lambda\sigma}(x)]}.
\end{eqnarray}
Note that we integrate the exponential oscillation form $e^{iS}$
rather than the exponential decay form $e^{-S}$. 
The latter is commonly and successfully used for Euclidean quantum field
theories whose action has quadratic form.
However, the action of general relativity is not quadratic but linear
in each of the variables.
Because of this fact, the action is not bound from the below and
the use of the exponential decay form runs into serious 
technical problems.
In addition, it unlikely contains the classical limit
since the value of the action for classical solutions is zero.
The former is used in other strategies in quantum gravity
\cite{JI3D,HO,CR,MR,FK,JI4D,BC,JB,sumover} even for Riemannian spacetime.
In the exponential oscillation form, large magnitudes of the action
less contribute to the path integral.
In order to examine possible consequences of the use of
the exponential decay form, we simply replace $\beta$ by $i\beta$
in the results of the use of the exponential oscillation form.

In terms of the path integral, define the expectation value of
an observable $X$ as follows.
\begin{eqnarray}
&&
\langle X\rangle_\Delta:=Z_\Delta^{-1}
\int d\zeta d\eta d\rho^4 X[\zeta,\eta,\rho]
e^{-iS_{\Delta}[\zeta,\eta,\rho]}.
\end{eqnarray}
In particular, we are interested in four quantities: 
$\langle\eta_\mu\rangle$, $\langle\zeta_\mu\rangle$, 
$\langle{\rm Tr}U_{\mu\nu}\rangle$ and $\langle S_\Delta\rangle$.
The first two are basic variables of the theory and are SU(2) dependent
quantities.
They should vanish.
This is because there is a symmetric structure along 
the SU(2) gauge trajectories.
It is known that the expactation value of the basic variables of lattice 
QCD vanishes unless an appropriate gauge is fixed.
This fact is often utilized to study gauge fixing ambiguities on lattice.
The same thing should happen here since the Gauss gauge constraint structure
of the present theory is common with the SU(2) version of QCD.

The third is (a smallest of) the Wilson loop.
It is SU(2) gauge independent and a good physical
observable for Gauss gauge invariant theories such as lattice QCD.
Here, it is possible that the expectation value of the Wilson loop vanishs
unless an appropriate gauge is fixed.
This would mean that the Gauss gauge invariance is not enough to be 
a physical observable for quantum gravity and a possible symmetric
structure along diffeomorphism gauge trajectories annihilates
the expectation value of the Wilson loop.
Quantum gravity is invariant not only under Gauss gauge transformations 
but also under spacetime diffeomorphisms.
The Wilson loops are not diffeomorphism invariant and hence cannot be
physical observables of quantum gravity.

The last one is the action of the theory.
It is invariant under SU(2) gauge transformations and diffeomorphisms and
the only known local physical observable in general relativity.
It should  vanish if the inverse coupling constant is brought to infinity
but diverge in the same limit if the path integral has the exponential decay
form.
The former would be an indication that the path integral successfully
eliminates the degrees of freedom constrained since the value of 
the action on the constraint surface must be identically zero.
The latter would be an indication that the path integral of 
exponential decay form cannot capture the classical solutions.
These are non-trivial checks for consistency.
These facts are examined in 3 dimensions.


\subsection{Character expansion}\label{subsec:character}

We compute the path integral as a character expansion and show that
it can be written as a sum over surface-like excitations in spacetime.
{}First we rewrite the path integral as follows.

\begin{eqnarray}
&&
Z_{\Delta}:=
\int d\zeta d\eta d\rho^4 
e^{-iS_{\Delta}[\zeta,\eta,\rho]}
\nonumber\\&&
=\int d\zeta d\eta d\rho^4 \prod_{x\in\Delta}
\prod_{(\mu\nu\lambda\sigma)}
e^{i2\beta\rho_\mu(x)\rho_\nu(x)
{\rm Tr}[\eta_\mu^\dagger(x)\eta_\nu(x)
U_{\lambda\sigma}(x)]}
e^{-i2\beta\rho_\mu(x)\rho_\nu(x)
{\rm Tr}[\eta_\nu^\dagger(x)\eta_\mu(x)
U_{\lambda\sigma}(x)]}.
\nonumber\\
\end{eqnarray}
Here, $(\mu\nu\lambda\sigma)$ denotes the even permutations of $(0123)$;
there are six terms, that is, $(0123)$, $(0231)$, $(0312)$, $(1203)$, 
$(3102)$, and $(2301)$.
Then use the formula (\ref{eq:eTr}) to expand to the characters
as follows.

\begin{eqnarray}
&&
Z_\Delta
=\int d\zeta d\eta d\rho^4 \prod_{x\in\Delta}
\prod_{(\mu\nu\lambda\sigma)}
\sum_{l_{\mu\nu},l_{\nu\mu}}
\Gamma_{\mu\nu}(\beta;\{l,\rho\})
\chi_{l_{\mu\nu}}(\eta_\mu^\dagger\eta_\nu U_{\lambda\sigma})
\chi_{l_{\nu\mu}}(\eta_\nu^\dagger\eta_\mu U_{\lambda\sigma}),
\label{eq:gxx}
\end{eqnarray}
with
\begin{eqnarray}
&&
\Gamma_{\mu\nu}(\beta;\{l,\rho\}):=
{(2l_{\mu\nu}+1)(2l_{\nu\mu}+1)(-1)^{l_{\mu\nu}-l_{\nu\mu}}\over
4\beta^2\rho_\mu^2\rho_\nu^2}
J_{2l_{\mu\nu}+1}(4\beta\rho_\mu\rho_\nu)
J_{2l_{\nu\mu}+1}(4\beta\rho_\mu\rho_\nu).
\nonumber\\
\end{eqnarray}
Here,
$\chi_j(U)$ is the character of SU(2) element $U$ in the spin-j 
representation and
$l_{\mu\nu}$ and $l_{\nu\mu}$ are spins  
taking values 0, $1\over2$, 1, ${3\over2}\cdots$.
$J_m(x)$ is the Bessel function of the first kind.
One of its definitions is given by (\ref{eq:J}).

{}From (\ref{eq:gxx}), 
one can understand how surface-like excitations emerge.
$U_{\lambda\sigma}(x)$ is a parallel transport along the four edges 
enclosing a face of the lattice.
The face is in the $\lambda\sigma$-plane and bases at $x$.
Let $f_{\lambda\sigma}(x)$ denote this face.
(We do not distinguish the orientations; 
$f_{\lambda\sigma}(x)\equiv f_{\sigma\lambda}(x)$.)
Associate the spin $l_{\mu\nu}$ of 
$\chi_{l_{\mu\nu}}(\eta_{\mu}^\dagger(x)\eta_{\nu}(x)U_{\lambda\sigma}(x))$
to the face $f_{\lambda\sigma}(x)$.
In the same way, $l_{\nu\mu}$ is also associated to the same face.
By integrating $\eta_{\mu}$ at $x$, one introduces an intertwiner 
to combine six spins $l_{\mu\nu}$ and $l_{\nu\mu}$ 
($\nu\ne\mu$ for given $\mu$) associated with $f_{\lambda\sigma}(x)$
($\lambda$ and $\sigma$ are such that $\epsilon^{\mu\nu\lambda\sigma}=1$).
By integrating $\zeta_\mu$ shared by six faces $f_{\mu\nu}(x)$ and 
$f_{\mu\nu}(x-\varepsilon^\nu)$ ($\nu\ne\mu$ for given $\mu$),
one introduces an intertwiner to combine the twelve spins associated with 
the six faces.
Therefore, after integrating all $\eta$'s and $\zeta$'s,
the spins and the intertwiners combining the spins are left.
Two spins are associated with a single face and two intertwiners are
associated with a single edge.
These facts are more clearly understood in the 3 dimensional
version of the model as examined in Sec. \ref{sec:3dim}.
The path integral is now written as a sum over spins and intertwiners
associated with faces and edges respectively.
A set of faces jointed by edges represent a surface-like object,
mathematically called 2-dimensional piece-wise linear cell complex 
\cite{JB}.
A surface-like excitation is a 2-dimensional piece-wise linear cell
complex with faces labeled by non-zero spins 
and with edges labeled by intertwiners.
A face with spin-0 means the absence of the face in a surface-like
excitation.

The integrations of $\eta$'s and $\zeta$'s can be explicitly done
by using the formulae (\ref{eq:D*}), (\ref{eq:DDtoD}) and (\ref{eq:DD}).
The resulting coefficiens contain the spins associated with faces and
other spins representing intertwiners associated with edges. 
The coefficients are expressed in terms of the so-called 3j-coefficients
(\ref{eq:3j}).
The integrations of $\rho$'s are difficult because of the presence of 
two $\rho$'s in the argument of the Bessel functions.
This difficulty is absent in the 3 dimensional version of the model.


\section{The model in 3-dimensions}\label{sec:3dim}

\subsection{The action and path integral}

{}From the action in 4 dimensions, it is straightforward
to write its 3 dimensional version.
We simply replace 
$l_p^{-2}d^4x\epsilon^{\sigma\lambda\mu\nu}e_\sigma^\dagger e_\lambda$
in the 4 dimensional action by
$l_p^{-1}d^3x\epsilon^{\lambda\mu\nu} e_\lambda$.
We compute the corresponding path integral on the lattice and show that
it is reduced to lattice BF theory in 3 dimensions in order to
claim that 3 dimensional version of the model exists and is a correct
quantum gravity model.

The action of general relativity in 3 dimensional Riemannian spacetime 
is defined as follows.
\begin{eqnarray}
&&
S_{R_{3}}:={1\over l_p}\int d^3x\epsilon^{\lambda\mu\nu}e_{\lambda i}
\left(2\partial_{[\mu}A_{\nu]}^i-\epsilon^i_{\ jk}A_\mu^jA_\nu^k\right).
\end{eqnarray}
Here, the Greek letter indices are $\{0,1,2\}$ for spacetime and
lower alphabet indices are $\{1,2,3\}$ for internal spacetime.

Let us count the number of degrees of freedom.
Out of 9 degrees of freedom of $e_\mu^i$ at each spacetime point, 
3 are for Lagrange multipliers
imposing two diffeomorphism and a Hamiltonian constraints.
Out of 9 degrees of freedom of $A_\mu^i$ at each spacetime point, 
3 are for Lagrange multipliers imposing 3 Gauss constraints.
Hence, 6 degrees of freedom in each are left.
They are cancelled by the 6 constraints and 6 gauge degrees of freedom
and one finds no local degrees of freedom consistently with general
relativity.
Note that in canonical theory one sees the vectors specifying 
a spacetime foliation multiplicative $e_\mu^i$.
Therefore, the integral over all foliations is taken into account
by the integrals of $e_\mu^i$.

The lattice action is defined as follows.
\begin{eqnarray}
&&
S_{\Delta_3}[\zeta,\eta,\rho]:=-\beta\sum_{x\in\Delta_3}
\epsilon^{\lambda\mu\nu}\rho_\lambda(x)
{\rm Tr}[\eta_\lambda(x)U_{\mu\nu}(x)].
\end{eqnarray}
Here, $\Delta_3$ denotes the 3 dimensional cubic lattice.
Note that the definitions of the variables are the same as for 
the 4 dimensional model, (\ref{eq:zeta}),(\ref{eq:eta}) and 
(\ref{eq:rho}), and the additional degrees of freedom $\eta_\mu^0$
cancel in the lattice action and do not play any role.
Then the lattice path integral is defined as follows.

\begin{eqnarray}
&&
Z_{\Delta_3}:=\int d\zeta d\eta d\rho^4\prod_{x\in\Delta_3}
e^{i\beta\epsilon^{\lambda\mu\nu}\rho_\lambda(x)
{\rm Tr}[\eta_\lambda(x)U_{\mu\nu}(x)]}
\nonumber\\&&
=\int d\zeta d\eta d\rho^4\prod_{x\in\Delta_3}
\prod_{(\lambda\mu\nu)=(012),\atop (120),(201)}
e^{i\beta\rho_\lambda{\rm Tr}[\eta_\lambda U_{\mu\nu}]}
e^{-i\beta\rho_\lambda{\rm Tr}[\eta_\lambda U_{\nu\mu}]}.
\end{eqnarray}
Then use the formula (\ref{eq:eTr}) to expand to the characters as follows.

\begin{eqnarray}
&&
Z_{\Delta_3}=\int d\zeta d\eta d\rho^4\prod_{x\in\Delta_3}
\prod_{(\lambda\mu\nu)=(012),\atop (120),(201)}
\sum_{l_{\mu\nu},l_{\nu\mu}}\Gamma_\lambda(\beta,\{l,\rho\})
\chi_{l_{\mu\nu}}(\eta_\lambda U_{\mu\nu})
\chi_{l_{\nu\mu}}(\eta_\lambda^\dagger U_{\mu\nu}),
\label{eq:Z3gxx}
\end{eqnarray}
with
\begin{eqnarray}
&&
\Gamma_\lambda(\beta,\{l,\rho\}):=
{(2l_{\mu\nu}+1)(2l_{\nu\mu}+1)(-1)^{l_{\mu\nu}-l_{\nu\mu}}
\over \beta^2\rho_\lambda^2}
J_{2l_{\mu\nu}+1}(2\beta\rho_\lambda)
J_{2l_{\nu\mu}+1}(2\beta\rho_\lambda).
\end{eqnarray}
Here, one can understand the emergence of surface-like excitations,
in other words, 2 dimensional piece-wise linear cell complexs
with faces labeled by non-zero spins and with edges labeled by intertwiners.
$U_{\mu\nu}$ is a parallel transport along the four edges enclosing
a face of the lattice.
The face is in the $\mu\nu$-plane and bases at $x$.
Let $f_{\mu\nu}(x)$ denote this face.
Associate the spin $l_{\mu\nu}$ of 
$\chi_{l_{\mu\nu}}(\eta_\lambda U_{\mu\nu})$
to the face $f_{\mu\nu}(x)$.
In the same way, $l_{\nu\mu}$ is also associated to the same face.
By integrating $\eta_\lambda$ at $x$, one introduces an intertwiner 
to combine two spins $l_{\mu\nu}$ and $l_{\nu\mu}$ associated to
$f_{\mu\nu}(x)$ ($\mu$ and $\nu$ are such that 
$\epsilon^{\lambda\mu\nu}=1$).
By integrating $\zeta_\mu$ shared by four faces $f_{\mu\nu}(x)$ 
and $f_{\mu\nu}(x-\varepsilon^\nu)$ ($\nu\ne\mu$ for given $\mu$),
one introduces an intertwiner to combine the eight spins associated with
the four faces.
Therefore, after integrating all $\eta$'s and $\zeta$'s, the spins and
the intertwiners combining the spins are left.
The path integral is now written as a sum over spins and intertwiners
representing surface-like excitations.

The integrations of $\eta$'s, $\zeta$'s and $\rho$'s can be explicitly
done by using the formulae (\ref{eq:D*}), (\ref{eq:DDtoD}) and 
(\ref{eq:DD}).
Integrate $\eta$'s and $\rho$'s as follows.

\begin{eqnarray}
&&
Z_{\Delta_3}=\int d\zeta\prod_{x\in\Delta_3}
\prod_{(\mu\nu)=(12),\atop (20),(01)}
\sum_{l_{\mu\nu}}\Delta_{\mu\nu}(\beta)
(2l_{\mu\nu}+1)\chi_{l_{\mu\nu}}(U_{\mu\nu}U_{\mu\nu}),
\label{eq:Z3UU}
\end{eqnarray}
with
\begin{eqnarray}
&&
\Delta_{\mu\nu}(\beta):=
\int_0^1 d\rho_\lambda^4
\left[{J_{2l_{\mu\nu}+1}(2\beta\rho_\lambda)\over\beta\rho_\lambda}\right]^2
=\beta^{-4}\int_0^{2\beta}dk k J_{2l_{\mu\nu}+1}(k)J_{2l_{\mu\nu}+1}(k)
\label{eq:delta}
\nonumber\\&&
\to\beta^{-4}\delta(0){\rm\ \ as\ \ }\beta\to\infty.
\end{eqnarray}

In the limit $\beta\to\infty$, $\Delta_{\mu\nu}$ loses the dependence on 
$l_{\mu\nu}$ and the path integral can be rewritten as follows.

\begin{eqnarray}
&&
\lim_{\beta\to\infty}Z_{\Delta_3}
=\int d\zeta\prod_{x\in\Delta_3}
\prod_{(\mu\nu)=(12),\atop (20),(01)}
\sum_{l_{\mu\nu}}(2l_{\mu\nu}+1)
[\chi_{l_{\mu\nu}}(U_{\mu\nu})+\chi_{l_{\mu\nu}}(-U_{\mu\nu})],
\end{eqnarray}
up to an overall multiplicative constant.
The first half is well known form for 3-dimensional lattice BF theory
and is meant that spacetime is flat $U_{\mu\nu}=1$.
In addition, we have another term also meaning that spacetime is flat but
$U_{\mu\nu}=-1$.
The presence of the additional term is due to the definition of the lattice
action.
Because of the definition of the lattice action, 
the path integral contains a form analogous to 
$\int dx e^{ix\sin\theta}=\delta(\theta)+\delta(\theta-\pi)$
instead of the well known form analogous to
$\int dx e^{ix\theta}=\delta(\theta)$.
We have understood that the path integral of the model in 3 dimensions
can be reduced to (a generalization of)
that of 3 dimensional lattice BF theory.


\subsection{Expectation values}\label{subsec:expectation}

We compute the expectation values of four quantities,
$\langle\eta_\mu\rangle$, $\langle\zeta_\mu\rangle$, 
$\langle{\rm Tr}U_{\mu\nu}\rangle$ and $\langle S_{\Delta_3}\rangle$,
and find all of them vanish.
However, their meanings are different as discussed below.
Remember that the path integral contains the integrations of
18 degrees of freedom per lattice point
(9 for $\zeta_\mu$ and 9 for $\eta_\mu$ and $\rho_\mu$).
If one translates them into the canonical theory,
3 of $\zeta_\mu$ and 3 of $\eta_\mu$ and $\rho_\mu$ are Lagrange
multipliers imposing the 6 constraints per lattice point.
The other degrees of freedom are 6 degrees of freedom constrained and
6 on the constraint surface.
The degrees of freedom constrained do not contribute to the path
integral.
When we compute the expectation values, the integrations over
the degrees of freedom on the constraint surface make non trivial
contributions.
All of the local degrees of freedom on the constraint surface are
gauge degrees of freedom in the 3 dimensional case.

$\langle\eta_\mu\rangle$ and $\langle\zeta_\mu\rangle$ are basic
variables of the theory.
They are SU(2) gauge dependent quantities.
It is known that the expectation value of basic variable of lattice
QCD vanishes unless an appropriate gauge is fixed because there is
a symmetric structure along the SU(2) gauge trajectories.
The same Gauss gauge constraint structure is also present in
general relativity in the present formulation.
Therefore, $\langle\eta_\mu\rangle$ and $\langle\zeta_\mu\rangle$
should vanish if the model is consistent.

Let us compute the expectation values of $\langle\eta_\mu\rangle$ and
$\langle\zeta_\mu\rangle$.
Insert $\eta_0(x)$ into (\ref{eq:Z3gxx}) and integrate $\eta_0(x)$.
This results in $l_{21}=l_{12}\pm{1\over2}$ associated with $f_{12}(x)$.
On the other hand, the integrations of $\eta_2(x)$,
$\eta_0(x-\varepsilon^2)$ and $\eta_2(x-\varepsilon^0)$ result in
$l_{10}=l_{01}$ associated with  $f_{01}(x)$, 
$l_{21}=l_{12}$ associated with $f_{12}(x-\varepsilon^2)$ and 
$l_{10}=l_{01}$ associated with $f_{01}(x-\varepsilon^0)$ respectively.
$f_{12}(x)$, $f_{01}(x)$, $f_{12}(x-\varepsilon^2)$ and 
$f_{01}(x-\varepsilon^0)$ share the edge where $\zeta_1(x)$ is defined.
Hence, the integration of $\zeta_1(x)$ introduces an intertwiner combining
the spins associated with the four faces.
However, no intertwiner consistent with these spins exists because the sum
of the 8 spins is a half integer.
Therefore, we conclude $\langle\eta_0\rangle=0$.
In the same way, we find $\langle\eta_1\rangle=\langle\eta_2\rangle=0$.

In order to compute $\langle\zeta_0\rangle$, insert $\zeta_0(x)$ into
(\ref{eq:Z3UU}) and integrate $\zeta_0(x)$.
The edge where $\zeta_0(x)$ is defined is shared by the four faces
$f_{01}(x)$, $f_{02}(x)$, $f_{01}(x-\varepsilon^1)$ and
$f_{02}(x-\varepsilon^2)$.
Since each of the four faces is associated with two equal spins
after the integrations of $\eta$'s,
the sum of the 8 spins associated with the four faces is an integer.
There is no intertwiner combining an integer and $1\over2$ to
produce the spin-0.
Therefore, we conclude $\langle\zeta_0\rangle=0$.
In the same way, we find $\langle\zeta_1\rangle=\langle\zeta_2\rangle=0$.

$\langle{\rm Tr}U_{\mu\nu}\rangle$ is the expectation value of
the Wilson loop and is an SU(2) gauge independent quantity.
It is a good physical observable for Gauss gauge invariant theories
such as lattice QCD.
However, the Gauss gauge invariance is not enough to be a physical
observable for quantum gravity.
Physical observables in quantum gravity must be invariant not only
under SU(2) gauge transformations but also under spacetime
diffeomorphisms.
The Wilson loop is not invariant under diffeomorphisms.
Therefore, it is possible that $\langle{\rm Tr}U_{\mu\nu}\rangle$
vanishes because of a symmetric structure along diffeomorphism
gauge trajectories unless an appropriate gauge is fixed.
If it is the case, the vanishing result can be understood as 
a consequence of the fact that quantum gravity is a spacetime 
diffeomorphism invariant theory.

The reasoning for $\langle\zeta_\mu\rangle$ to vanish also holds
for $\langle{\rm Tr}U_{\mu\nu}\rangle$ to vanish.
Since every face is associated with two equal spins 
after the integrations of $\eta$'s,
there is no way of combining the 8 spins associated with the four faces
and spin $1\over2$ of ${\rm Tr}U_{\mu\nu}$ to produce the spin-0.
Therefore, we find $\langle{\rm Tr}U_{\mu\nu}\rangle=0$.
The fact that every face is associated with two spins is a notable
difference of the model from lattice QCD.
One of the reasons for this fact is the presence of the variable $e_\mu^i$
in addition to the connection variable $A_\mu^i$ in the action.
Note that $\langle{\rm Tr}U_{\mu\nu}\rangle=0$ does not contradict to
the fact that the same quantity evaluated with flat connection is not zero
but one since this loop can be contracted to a point.
This fact is restored by the computation of the quantity with a gauge
fixing corresponding to either $U_{\mu\nu}=1$ or $-1$.
The vanishing result is due  to the symmetric structure consisting of
$U_{\mu\nu}=1$ and $-1$ on the constraint surface.

$\langle S_{\Delta_3}\rangle$ is the expectation value of the action.
It is invariant under SU(2) gauge transformations and 
spacetime diffeomorphisms.
The value of the action on the constraint surface is identically zero.
Therefore, $\langle S_{\Delta_3}\rangle$ must vanish if the path integral
successfully eliminates the contributions from the degrees of freedom
constrained.
This must be the case if the path integral adopts the exponential
oscillation form with $\beta\to\infty$.
On the other hand, if the path integral adopts the exponential decay form,
then the dominant contribution comes from the value of the action far
away from zero and hence $\langle S_{\Delta_3}\rangle$ unlikely vanishes.
This fact can be checked by replacing $\beta$ in the path integral
by $i\beta$.

In order to compute $\langle S_{\Delta_3}\rangle$,
take a derivative of (\ref{eq:Z3UU}) with respect to $\beta$
(with $i$ multiplied)
instead of inserting $S_{\Delta_3}$ into (\ref{eq:Z3gxx}).
This procedure results in the following quantity inserted in 
the path integral.
\begin{eqnarray}
&&
\sum_{x\in\Delta}\sum_{(\mu\nu)=(12),\atop (20),(01)}
{i{d\over d\beta}\Delta_{\mu\nu}(\beta) \over\Delta_{\mu\nu}(\beta)}
\end{eqnarray}
Since there is no particular place or direction in
purely empty spacetime, the every term must have equal contrubution and we
compute one of the terms as follows.
\begin{eqnarray}
&&
{i{d\over d\beta}\Delta_{\mu\nu}(\beta) \over\Delta_{\mu\nu}(\beta)}=
-{4i\over\beta}+{4i[J_{2l_{\mu\nu}+1}(2\beta)]^2\over\beta^3\Delta_{\mu\nu}}\to
-{4i\over\beta}\left(1-{1\over c\pi}\cos^2[2\beta-{\pi\over2}(2l_{\mu\nu}+1)
-{\pi\over4}]\right),
\label{eq:zero}
\nonumber\\
\end{eqnarray}
as $\beta$ goes to infinity.
Here, we have used the asymptotic formula
$J_m(x)\to\sqrt{2\over\pi x}\cos(x-{m\pi\over2}-{\pi\over4})$ for large $x$
and the fact that $\Delta_{\mu\nu}$ goes to
$\beta^{-4}\delta(0)$ or more precisely $c\beta^{-3}$ as $\beta\to\infty$,
where $c$ is a constant. 
We do not need the detailed value of $c$.
This fact can be understood by rescaling $\beta$ in (\ref{eq:delta}).
{}From (\ref{eq:zero}), we conclude $\langle S_{\Delta_3}\rangle\to 0$
as $\beta\to\infty$.
Note that if one replaces $\beta$ by $i\beta$ in the path integral, then
one finds (\ref{eq:zero}) with the cosine replaced by the hyperbolic cosine.
In this case, $\langle S_{\Delta_3}\rangle$ diverges instead of converging
to zero as $\beta\to\infty$.


\section{Conclusion}\label{sec:conclusion}

In the present work, we constructed a surface-theoretic lattice quantum
gravity model in 4 dimensional Riemannian spacetime based on the SU(2)
Ashtekar formulation of general relativity.
We introduced a version of the action of general relativity and defined
its lattice version on a fixed hyper cubic lattice.
We introduced  a dimensionless ``(inverse) coupling'' constant
so that the magnitude of the action is finite per lattice point.
We defined a path integral whose integrand has the exponential oscillation
form so that it eliminates the degrees of freedom constrained.
The finiteness of the action with finite lattice allowed the path integral
without gauge fixing finite.
We expanded the path integral in the SU(2) characters and showed that
the path integral can be written as a sum over surface-like excitations
in spacetime.
We showed that the 3 dimensional version of the model exists and 
its path integral is reduced to that of 3-dimensional lattice BF theory.
Therefore, we considered the model as a 4 dimensional generalization of
the Ponzano-Regge model with local degrees of freedom.
We examined the expectation values of two basic variables of the theory,
the Wilson loop and the action of the model in 3 dimensions
and showed all of them vanish.
We discussed the meaning of each of them and understood that the model
has a chance of capturing 
the physical degrees of freedome of general relativity.

\appendix

\section{Useful formulae}\label{sec:useful}

In the present work, we have used the following mathematical formulae.
\begin{eqnarray}
&&
e^{ix{1\over2}{\rm Tr}U}=\sum_j 2{2j+1\over x}i^{2j}J_{2j+1}(x)\chi_j(U),
\label{eq:eTr}
\\&&
J_m(x)={1\over2\pi}\int_{-\pi}^{\pi}d\theta e^{ix\sin\theta-im\theta}.
\label{eq:J}
\end{eqnarray}
The first formula can be easily proved and the second is a definition 
of the Bessel function of the first kind.
$\chi_j(U)$ is the character of SU(2) element $U$ in the spin-j 
representation and $j$ is a spin 
taking values 0, $1\over2$, 1, ${3\over2}\cdots$.

\begin{eqnarray}
&&(-1)^{n-m}D_{nm}^{(j)}(U)=D_{-n,-m}^{(j)*}(U),
\label{eq:D*}
\\&&
D_{n_1m_1}^{(j_1)}(U)D_{n_2m_2}^{(j_2)}(U)=
\sum_{j,n,m}(2j+1)
\left(\matrix{j_1&j_2&j\cr n_1&n_2&n}\right)
\left(\matrix{j_1&j_2&j\cr m_1&m_2&m}\right)
D_{nm}^{(j)*}(U),{\ \ }
\label{eq:DDtoD}
\\&&
\int dU D_{mn}^{(i)}(U)D_{m'n'}^{(j)*}(U)=
{1\over 2j+1}\delta_{ij}\delta_{mm'}\delta_{nn'},
\label{eq:DD}
\end{eqnarray}
$D_{mn}^{(j)}(U)$ is the spin-j representation matrix of SU(2) 
element $U$
and $m$ and $n$ run from $-j$ through $j$ with the increment 1.
$\left(\matrix{j_1&j_2&j\cr m_1&m_2&m}\right)$ 
is the so-called 3j-coefficient and 
defined by 
\begin{eqnarray}
&&
\left(\matrix{j_1&j_2&j\cr m_1&m_2&m}\right):=
{(-1)^{j_1-j_2-m}\over\sqrt{2j+1}}
\langle j_1m_1;j_2m_2|j,-m\rangle,
\label{eq:3j}
\end{eqnarray}
with the Clebsch-Gordan coefficient $\langle j_1m_1;j_2m_2|jm\rangle$.
The asterisks mean the complex conjugate and the sum of $j$ 
is taken over $|j_1-j_2|$ through $j_1+j_2$ and the sums of $n$ and $m$
over $-j$ through $j$.




\begin{thebibliography}{Refer 999}

\bibitem{JI3D} J. Iwasaki,
``{\it A reformulation of the Ponzano-Regge quantum gravity
model in terms of surfaces},"
gr-qc 9410010;
``{\it A definition of the Ponzano-Regge quantum gravity
model in terms of surfaces},"
gr-qc 9505043, J. Math. Phys. 36 (1995) 6288.

\bibitem{HO} H. Ooguri,
``{Topological lattice models in four dimensions},"
Mod. Phys. Lett. A7 (1992) 2799-2810.

\bibitem{CR} M. Reisenberger, C. Rovelli,
``{\it Sum-over-surface form of loop quantum gravity},"
gr-qc 9612035, Phys. Rev. D 56 (1997) 3490;
C. Rovelli,
``{\it The projector on physical states in loop quantum gravity,}"
gr-qc 9806121, Phys. Rev. D 59 (1999) 104015;
``{\it The century of the incomplete revolution: 
searching for general relativistic quantum field theory},''
hep-th 9910131;



\bibitem{review} 
R. Gambini, J. Pullin, 
``{\it Loops, Knots, Gauge Theory and Quantum Gravity},''
Cambridge University Press, Cambridge, 1996;
C. Rovelli, ``{\it Loop quantum gravity}," gr-qc 9710008,
Living Reviews, volume 1.

\bibitem{AA} A. Ashtekar,
``{\it Lectures on non-perturbative canonical gravity},''
World Scientific Publishing, Singapore, 1991.


\bibitem{MR} M. Reisenberger,
``{\it Worldsheet formulations of gauge theories and gravity},"
gr-qc 9412035; 
``{\it A left-handed simplicial action for euclidean general relativity},"
gr-qc 9609002, Class. Quant. Grav. 14 (1997) 1753-1770;
``{\it A lattice worldsheet sum for 4d Euclidean general relativity},"
gr-qc 9711052.

\bibitem{FK} L. Freidel, K. Krasnov, 
``{Spin foam models and the classical action principle}," 
hep-th 9807092, Adv. Theor. Phys. 2 (1998) 1221-1285.

\bibitem{JI4D} J. Iwasaki,
``{\it A surface theoretic model of quantum gravity}," 
gr-qc 9903112, in The Proceedings of The 3rd Mexican School on Gravitation
and Mathematical Physics: Black holes, Classical and Quantum,
Nov. 15-20 (1998), Mazatran, Sinaloa, Mexico, Science Network
Publishing.

\bibitem{JP} J. Plebanski, J. Math. Phys. 18 (1977) 2511.

\bibitem{BC} J. Barrett, L. Crane,
 ``{\it Relativistic spin networks and quantum gravity},"
gr-qc 9709028, J. Math. Phys. 39 (1998) 3296.

\bibitem{JB} J. Baez,
``{\it Spin foam models}," gr-qc 9709052, 
Class. Quant. Grav. 15 (1998) 1827-1858;
``{\it An introduction to spin-foam models of quantum gravity and 
BF theory}," gr-qc 9905087;
``{\it Spin form perturbation theory},''
gr-qc 9910050.

\bibitem{MS} F. Markopoulou, L. Smolin,
``{\it Causal evolution of spin networks}," gr-qc 9702025,
Nucl. Phys. B 508 (1997) 409;
{}F. Markopoulou, ``{\it Dual formulation of spin network evolution},"
gr-qc 9704013.

\bibitem{SSJ} J. Samuel, Pramana J. Phys. 28 (1987) L429;
T. Jacobson, L. Smolin, Class. Quant. Grav. 5 (1988) 583. 


\bibitem{KW} K. Wilson, Phys. Rev. D 10 (1974) 2445.

\bibitem{sumover} M. Reisenberger, C. Rovelli,
``{\it Spin foams as Feynman diagrams}.'' gr-qc 0002083;
``{\it Spacetime as Feynman diagrams: the connection formulation},'' 
gr-qc 0002095;
R. DePietri, L. Freidel, K. Krasnov, C. Rovelli,
``{\it Barrett-Crane model from a Boulatov-Ooguri field theory 
over a homogeneous space},'' hep-th 9907154.


\bibitem{AL} A. Ashtekar, J. Lewandowski,
J. Math. Phys. 36 (1995) 2170;
A. Ashtekar, J. Lewandowski, D. Marolf, J. Mour\~ao, T. Thiemann,
J. Math. Phys 36 (1995) 6456.


\end{thebibliography}
\end{document}